# Cache Bypassing and Checkpointing to Circumvent Data Security Attacks on STTRAM

Nitin Rathi, Asmit De, *Helia Naeimi and Swaroop Ghosh
Computer Science and Engineering, University of South Florida
*Intel Labs, Santa Clara, CA
{nitinr, swaroopghosh, asmitde}@mail.usf.edu, helia.naeimi@intel.com

*Abstract- Spin-Transfer Torque RAM (STTRAM) is promising for cache applications. However, it brings new data security issues that were absent in volatile memory counterparts such as Static RAM (SRAM) and embedded Dynamic RAM (eDRAM). This is primarily due to the fundamental dependency of this memory technology on ambient parameters such as magnetic field and temperature that can be exploited to tamper with the stored data. In this paper we propose three techniques to enable error free computation without stalling the system, (a) stalling where the system is halted during attack; (b) cache bypass during gradually ramping attack where the last level cache (LLC) is bypassed and the upper level caches interact directly with the main memory; and, (c) checkpointing along with bypass during sudden attack where the processor states are saved periodically and the LLC is written back at regular intervals. During attack the system goes back to the last checkpoint and the computation continues with bypassed cache. We performed simulation for different duration and frequency of attack on SPLASH benchmark suite and the results show an average of 8% degradation in IPC for a one-time attack lasting for 50% of the execution time. The energy overhead is 2% for an attack lasting for the entire duration of execution.*

## I. Introduction

Spin-Transfer Torque RAM (STTRAM) [1] is promising for Last Level Cache (LLC) due to numerous benefits such as high-density, non-volatility, high-speed, low-power and CMOS compatibility. Fig. 1 shows the STTRAM cell schematic with Magnetic Tunnel Junction (MTJ) as the storage element. The MTJ contains a free and a pinned magnetic layer. The resistance of the MTJ stack is high (low) if free layer magnetic orientation is anti-parallel (parallel) compared to the fixed layer. The MTJ can be toggled from parallel to anti-parallel (or vice versa) by injecting current from source-line to bitline (or vice versa). The data in MTJ is stored in the form of magnetization. The data stored is '1' if the free layer magnetization is anti-parallel to fixed layer magnetization and '0' if they are parallel. The read/write latency of MTJ depends on the size of the device, current passing through the layers as well as on process variation.

The free layer of MTJ flips under the influence of external magnetic field which can be exploited by the adversary to launch magnetic attacks. The magnetic field produced by a horseshoe magnet can be used to flip the weak bits in a STTRAM memory array resulting in scrambling of stored data [2]. The switching process and retention time of MTJ is highly sensitive to temperature. The retention time of MTJ reduces at high temperature and the bits start to flip randomly under the influence of thermal noise. Therefore, magnetic field and temperature can be exploited by the adversary to scramble the data in LLC to launch denial of service (DoS) attack or simply increase the miss rate affecting the overall performance of the system. The existing countermeasures to mitigate magnetic attack include variable strength error correcting code (ECC) and forced retention [2]. The strength of the ECC is increased (1bit/2bit/4bit/8bit) depending on the magnitude of the attack. The ECC design is modular and during normal operation the unused ECC modules are power gated to reduce energy. Although effective ECC introduces significant design overhead. The effect of temperature on the read/write current, latency and bit error rate is presented in [3] however mitigation technique is not provided.

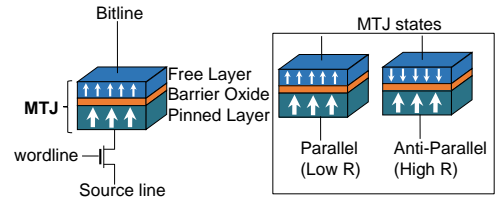

Fig. 1 Schematic of STTRAM bitcell showing MTJ.

In this paper we consider two types of magnetic attack on STTRAM LLC. In the first case the strength of the attack ramps up gradually and in the second case we assume a sudden rise in attack strength. The gradual ramping attack is applicable to attacks launched by bringing a permanent magnet or electromagnet in close proximity to the memory manually. The sudden attack applies to scenarios where the adversary has precise control over parameters such as magnetic field strength and distance from the chip. If the detected attack strength is beyond the threshold where the functional bits fail it is assumed to be a sudden attack. The attack signal is generated by the sensors [2] distributed across the memory array. The sensors are composed of MTJ cells which are designed to be less robust than the actual functional bits. The sensor MTJs [2] can sense both gradually ramping attack as well as sudden attack through fail rate. Based on the sensor input we propose a suite of techniques to deal with the attack. A simple stalling is proposed where the execution of instructions is stalled during the ramping attack and the execution resumes from the same state after the attack is removed. Write back of dirty data is performed before stalling to make the processor state updated. The LLC is invalidated before resuming the execution to avoid reading of any corrupted data. Cache bypassing is proposed to continue error-free computation during the ramping attack (Fig. 2(a)). The attack sensors detect the attack ahead of time and the system is prepared to enable bypassing. The system needs to write back the dirty data in case of write-back policy to save the modifications made before the attack. Write back consumes some clock cycles before the system can continue with the LLC bypassing. This step is shown in the figure as preparing for bypass. After the write-back the bypassing is enabled and the system runs at lower performance due to long memory latency. In ramping attack the sensors sense the attack ahead of time to perform the write back but in case of a sudden attack there is no opportunity to write back the dirty data as the functional bits start failing instantly (Fig. 2(b)). We propose checkpointing technique where CPU register values and program counter (PC) are saved in a non-volatile memory and write back is performed on all cache levels. In the event of an attack the processor states are loaded with the last saved checkpointed data and the pipeline is flushed. The instructions executed between the last checkpoint and detection of attack are re-executed (Fig. 2(b)). The LLC is bypassed during the attack to prevent functional failures. Once the attack is removed the LLC is invalidated and the bypass signal is de-asserted. The system continues to perform checkpointing at regular intervals.

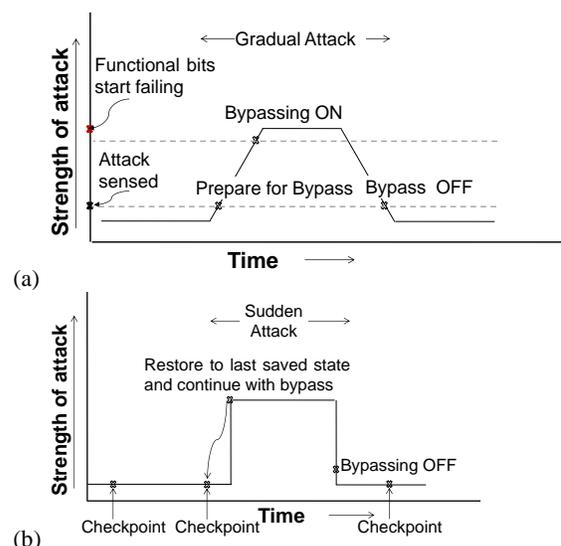

Fig. 2 Proposed mitigation techniques for, (a) gradually ramping attack; and, (b) sudden attack. Both techniques can be combined together for greater resilience towards attack.

*Related work and contributions:* Cache bypassing has been proposed previously to increase the performance and effective capacity of LLC without incurring power/area costs of a larger sized cache. The idea is to bypass the blocks which may pollute the cache [4][5]. A significant number of items referenced in a program are accessed very rarely and when they are fetched in cache they evict other cache blocks. In such cases not only it nullifies the benefit in placing those items in cache but it also incurs eviction overhead of blocks (which may be one of the frequently accessed blocks) to make way for these not so frequently accessed blocks. Furthermore, since the data is fetched from the main memory in block sizes (512KB/1024KB), fetching one word leads to the eviction of entire cache line. In such scenarios the best option is to bypass the cache and directly send the requested word to CPU. Intel's i860 processor provides support for cache bypassing [6]. A load instruction PFLD (pipelined floating-point load) is provided to bypass the LLC to avoid cache pollution. Cache bypassing is proposed for STTRAM LLC since the latency of write operations is 2X higher than read operations which may obstruct other cache accesses on a multi-core system running multiple processes. Therefore, other accesses can be forwarded to the main memory or upper level caches [7]. Similarly, the reusability of cache blocks is very low in GPGPU applications where cache bypassing results is higher performance [8].

Note that the existing bypass techniques are one-way, i.e., they bypass the data coming from main memory to LLC. The data coming from CPU to LLC is not bypassed. Therefore these techniques cannot be extended for data security. Other factors which require a new bypass architecture for security are as follows: (a) Bypassing of both ways is required because the LLC under attack is not safe to read as well as write new data; (b) Bypassing needs to be dynamically enabled and disabled depending on attack signal from sensors; and, (c) Before starting bypassing the dirty blocks in LLC should be written back to memory and after the bypass ends the LLC should be invalidated.

In this paper we propose a LLC bypass architecture which meets the above requirements and skips all the traffic to LLC, both from upper level cache and main memory seamlessly under attack. Although promising the LLC bypassing cannot handle sudden attacks. This is true because entering the bypass mode takes several cycles which is unavailable during sudden attack. We propose checkpointing to ensure that the system saves its state periodically. This enables us to enter the bypass mode with previously saved state during attack.

System-level checkpointing is a mechanism used in modern systems to provide recovery in case of sudden power failure [9]. Micro-architectural checkpointing is also proposed for system recovery from transient faults [10]. The basic approach is to perform computations in epochs during which the underlying hardware is checked for errors, if any fault is detected the results of that epoch is discarded and the system is restored to last known good state. During an epoch the results are held in a speculative state and get committed at the time of checkpointing. System-level state checkpointing has been employed to improve the performance of reorder buffer (ROB) in terms of handling exceptions [11]. Application level self-checkpointing techniques also exists [12]. The proposed checkpointing mechanism in this paper has been adopted from [10]. Since checkpointing is associated with IPC and energy overhead, the frequency of checkpointing could be tuned according to attack frequency. Initially, the checkpointing can be performed at larger intervals and on detection of attack the frequency of checkpointing can be increased. Note that the proposed bypassing technique can also be extended for thermal attack (using temperature sensors) or even for situations when the temperature of the LLC increases due to ambient conditions.

*To the best of our knowledge this is the first effort towards protecting data security from non-invasive tampering on LLC using micro-architectural features.* Although we focus on the LLC the proposed techniques can also be extended for non-volatile main memory. In summary, we make following contributions in this paper:

- We propose CPU stalling technique to handle ramping attack with least design complexity.
- We propose a novel dynamic LLC bypassing technique which exploits the existing design features to enable safe computing seamlessly under ramping attack.
- The LLC bypass incurs a maximum IPC overhead of 40% when attack persists throughout the program execution and 15% when attack occurs only once for half the duration.
- We propose checkpointing with LLC bypass to handle sudden attacks on LLC.

The rest of the paper is organized as follows. In Section II, we describe the STTRAM attack models. The bypassing architecture is described in Section III. The simulation results are presented in Section IV. Conclusions are drawn in Section V.

## II. STTRAM Attack Models

The flipping of MTJ is very sensitive to magnetic field and temperature. The dynamics of the MTJ free layer is described by the LLG equation [13].

$$\frac{\partial \vec{m}}{\partial t} = -\gamma \vec{m} \times H_{eff} - \alpha\gamma\vec{m} \times \vec{m} \times H_{eff} + \underbrace{\frac{I_s \hbar G(\psi)}{2e} \vec{m} \times (\vec{m} \times \vec{e_p})}_{STT}$$

Where $\vec{m}$ is unit vectors representing local magnetic moment, $\alpha$ represents the Gilbert's damping parameter, $\gamma$ is gyromagnetic ratio, $I_s$ is spin current, $G(\psi)$ is the transmission co-efficient, $\hbar$ is reduced planck's constant, e is charge on electron and $\vec{e_p}$ is the unit vector along fixed layer magnetization. In the above expression $\vec{H_{eff}}$ is effective field. The adversary can place an external AC/DC magnetic field to alter the $\vec{H_{eff}}$ parameter resulting in uneven flipping during read/write operation [2]. The magnetic field can also be used to force the free layer magnetization to flip under retention [2]. Heating can also be used to affect the magnetization reversal of MTJ during read/write operation [2]. The retention time of MTJ is exponentially related to thermal barrier ($\Delta$) by $t = C \times e^{k\Delta}$, where $C$ and $k$ are fitting constants. The thermal barrier is inversely proportional to absolute temperature (T) by $\Delta = \frac{E}{k_B T}$, where E is the energy barrier and $k_B$ is the Boltzmann's constant. Therefore the retention time at high

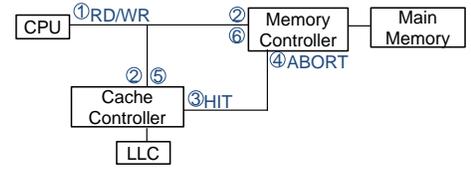

① CPU issues a read or write request. Requested tag misses in L1 and L2.
② Both cache controller and memory controller receive request simultaneously.
③ Cache controller searches tag array for requested tag and asserts HIT signal if match is found.
④ Asserted HIT signal is connected to ABORT signal of memory controller, so if HIT=1 memory controller aborts, or else continues to serve the CPU request.
⑤ If the data is found in cache the cache controller sends the data to CPU.
⑥ Else the data is received from main memory temperature reduces and the bits start to randomly flip resulting in scrambling of sensitive user data. Heating the memory array along with external magnetic field may result in faster switching of MTJs. Both heating and magnetic attack can be carried out even when the system is OFF. But such attacks will not affect the computation as the cache is invalidated on startup. Therefore, we focus on active attacks, i.e. when the system is operational.

Fig. 3 Look aside cache architecture.

## III. Prevention Techniques

In this section we present three countermeasures to protect against data security attacks on STTRAM LLC.

*A. System Assumptions*

We assume following features in the system for analysis:

*Attack sensors:* We assume that the attack signal can be asserted by the sensors [2] before the actual bits are affected. Depending on the sensor error rate a signal can also be asserted to indicate whether the attack is gradual or sudden.

*Exclusive LLC:* Exclusive cache do not store the copy of data already present in upper level caches. The data is guaranteed to be present in one of the caches. The advantage of exclusive cache is that they store more data. Another advantage is that in case of bypassing we do not need to worry about maintaining the inclusive property. Note that bypassing in inclusive LLC can also be implemented by storing the tags of the bypassed blocks in a buffer to maintain the inclusive property. The algorithm which checks the inclusiveness will compare the tags in all cache levels and raise an exception in case a tag is present in upper level cache and its copy is missing in LLC [4].

*Write-no-allocate:* In write-no-allocate policy the write misses are not loaded in the cache and the data is directly written to the main memory. In this policy only the reads are cached. We employ this policy to avoid unnecessary update of cache during write miss in bypass mode.

*Look-aside cache*: Look aside cache architecture is a system where the cache is located on the processor bus in parallel with the main memory controller (Fig. 3). This design enables both the cache controller and memory controller to service

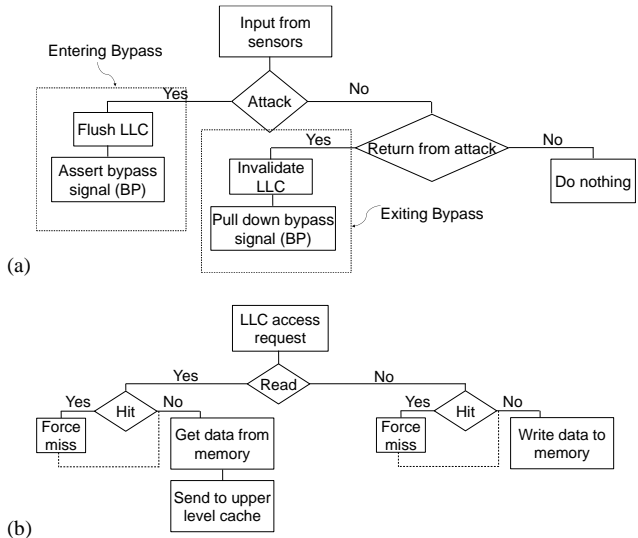

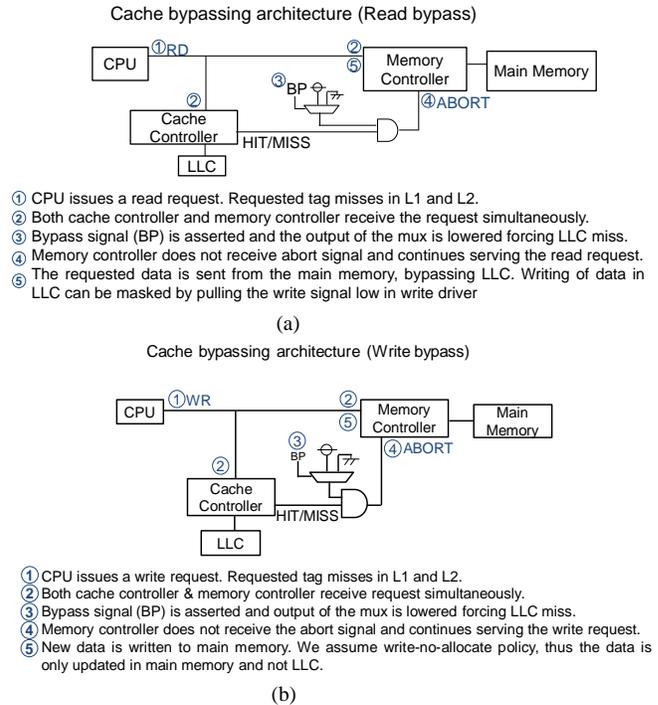

Fig. 4(a) Control flow of activating and deactivating bypassing, and (b) read, write request processing during bypassing.

CPU read and write requests simultaneously. If a cache miss occurs, then the request is completed by the memory controller. Fig. 3 explains the read/write operation in a look-aside cache architecture. The CPU issues a read/write request and if the requested tag is found in upper level caches (L1, L2) then it is serviced by them. If a miss occurs in upper level caches (step 1), then the request is simultaneously sent to both LLC cache controller and main memory controller as they are connected to the processor bus in parallel (step 2). The cache controller searches the requested tag in LLC and asserts the HIT signal if it is found (step 3). The assertion of HIT signal sends an ABORT signal to the memory controller informing that the tag is found in LLC and the memory controller should abort searching in main memory. The corresponding data is then sent to the CPU from the LLC (step 5). If the tag is not found in LLC then the HIT and ABORT signals stay de-asserted and the data is fetched by the memory controller. The corresponding data is sent to both CPU and LLC from the main memory (step 6). Therefore, the memory access time is reduced during LLC miss compared to traditional *look-through* cache.

### B. Stalling

The simplest and robust solution is to stall the CPU and wait till the attack is over. If the cache implements write-back policy, then the dirty data is written back to the main memory to save the system state on detection of the attack (for gradually ramping attack) and the CPU is stalled. After the attack is over, the entire LLC is invalidated and the computation starts from the last saved state. The processor's register contents will remain intact and the computation can begin from the state it was halted. This technique is better than shutting down the entire system because the processor states remains intact and the computation can instantly start after the attack is over. For the user, the machine will appear to be stuck during the attack however the user is not required to reboot the system. Although simple, this technique will not work for sudden attack since the dirty data will be corrupted (or become untrustworthy). For such scenarios the processor has to be restarted after the attack and the applications can restore the states if application level checkpointing [14-15] is implemented (which is typically the case for common applications such as Microsoft word, powerpoint, firefox). Both of the above approaches will disable computations during attack and result in power loss. The attacker can also exploit these features to drain the battery of the system.

### C. Cache Bypass

Cache bypassing enhances the user experience as the computation continues with affordable IPC degradation. We show the necessary steps needed to prepare for bypassing, continue bypassing and exit bypassing (Fig. 4(a)). If the sensors indicate a weak attack the LLC is flushed by copying the dirty data and a bypass signal (BP) is asserted. In absence of attack, if the bypass signal is still asserted (indicating the end of attack), the entire LLC is invalidated and the BP signal is de-asserted. Otherwise, no extra steps are needed. In the following paragraphs we explain various stages of bypassing.

*Preparing for bypassing (Fig. 4(a)):* If the sensors indicate an attack the dirty data in LLC is copied to the main memory by asserting the FLUSH signal [16] in the cache controller to ensure correctness. Note that this is possible since the sensors can sense the attack ~100us ahead [2]. The FLUSH signal writes back the dirty blocks and invalidates all the cache lines after the write-back. In cases where the LLC has write-through policy this step is not necessary as the copy of data is immediately written back to main memory. The BP is asserted to indicate the cache controller to bypass the subsequent requests to the main memory.

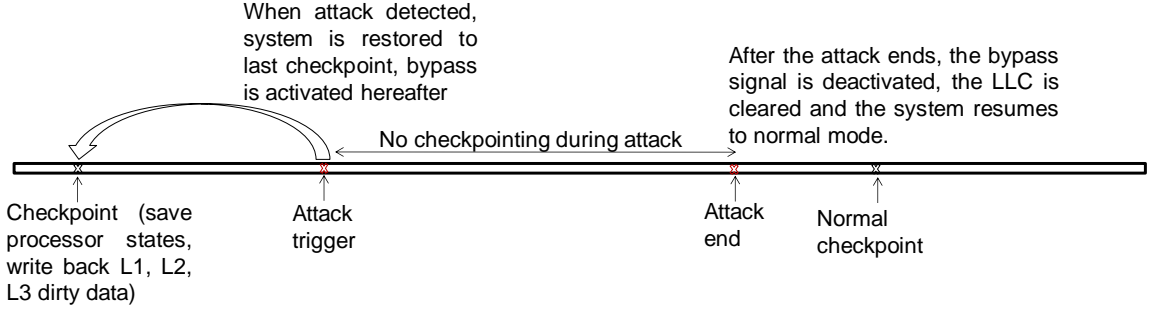

Fig. 6 Cache bypass architecture with checkpointing.

*Bypassing mode (Fig. 4(b)):* There are four scenarios when the data can leave or enter the LLC namely, read hit, read miss, write hit and write miss. The read hits are forcibly converted to read misses so that the data is read from the main memory instead of cache. Read misses are served normally by sending the data from main memory. Write hits are also forcibly converted to write misses and the data is written only to main memory (write –no-allocate). In case of write misses the main memory is updated with the new data. During the attack LLC data should not be used for computation or stored anywhere (upper level caches, main memory). However, new data may be written in the LLC during attack (to simplify design complexity) which will result in energy overhead but will not cause functional failure. In the following paragraphs we explain the implementation of bypass during various cache accesses:

*(i) LLC read hit (Fig. 5(a)):* If the address requested by the CPU is not found in the upper level caches the request is forwarded to the LLC. If tag match happens in LLC and the corresponding valid bit is set, then the data is sent to CPU. In case of tag miss or invalid data in LLC the cycle is completed by the main memory as described earlier. To enable bypass we add a multiplexer and an AND gate to force a LLC miss. Therefore, even if the data is present in LLC the cache controller is tricked to send a miss signal and the data is fetched from main memory. The data may be overwritten in LLC with new data incurring energy overhead.

*ii) LLC read miss (Fig. 5(a)):* If the address requested by the CPU is not found in any level of cache then the request is forwarded to memory controller and the data is read from main memory. A copy of the data is also placed in LLC. In the proposed architecture all the read requests are forced to be a LLC miss and each time the data is taken from the main memory if it is not present in upper level caches.

*iii) LLC write hit (Fig. 5(b)):* If the write cycle issued by the CPU matches the tag in LLC then the corresponding data is updated. But during bypass all write requests on LLC are forced to be a miss and the CPU writes in main memory directly (assuming write-no-allocate policy). Thus, the LLC under attack is not updated with the new data.

*iv) LLC write miss (Fig. 5(b)):* In case of LLC write miss when the requested address is not found the writes are automatically forwarded to the main memory. During bypass, all write requests are forced to be a miss and the main memory is always updated with the new data.

*Exiting bypass mode (Fig. 4(a)):* If the attack ends and the system was not under attack then no action is needed. If the system was in bypass mode, then we invalidate the entire LLC after attack since the data cannot be trusted. After the bypass signal is de-asserted the subsequent requests are serviced by the LLC. A hardware interrupt is forced to stall the CPU and prevent updating of LLC during the flush and invalidate operations.

## C. Checkpointing

We leverage the system-level checkpointing to mitigate the sudden attacks. Fig. 6 illustrates the high-level timeline of execution of events performed during a sudden attack. The CPU register values and PC are saved in a non-volatile memory and write back is performed on all cache levels during checkpointing event. When attack is sensed the system is restored to the last saved checkpoint and the bypass signal is asserted. The system continues to perform with the LLC bypass and the checkpointing is disabled to avoid write back of stale LLC data. After the attack ends the bypass signal is deactivated and the LLC is invalidated. The system continues to perform normally and the checkpointing is resumed. The LLC write buffers are prevented from committing the data to main memory between two checkpoints. If the write buffers are not masked, it may cause functional failures when the system restores to a last checkpoint. The instructions executed after the checkpoint should be discarded in case of roll back.

Table 1: Processor Configuration

| Processor | Alpha, O3, 4 cores, 2GHz, 8-way issue |
|---|---|
| SRAM L1-Cache | Private, Icache=16KB, Dcache=16KB, LRU, 64B cache-line, 2 cycle read/write latency, write back. |
| SRAM L2 Cache | Private, 256KB, 8-way, LRU, 64B cache-line, 8 cycle read/write latency, write-back. |
| LLC Cache | Shared, 8MB, 4 banks, 8 ways, 8-entry write buffer per bank, 17cycle read latency, 34 cycle write latency, write-back, write no-allocate. |

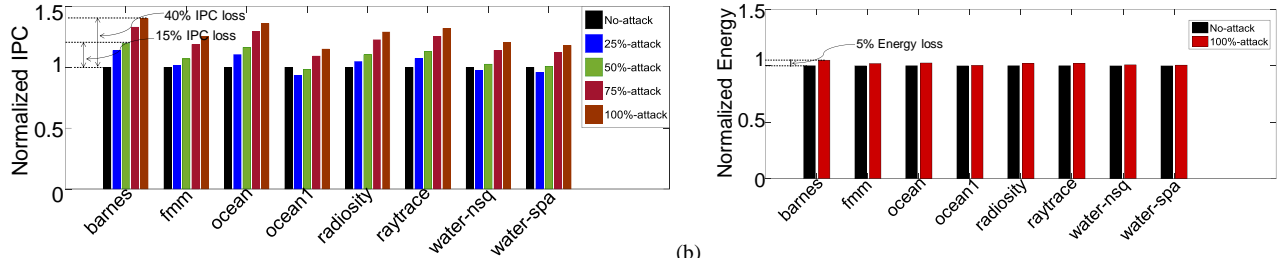

Fig. 7 (a) Normalized IPC, and; (b) normalized energy for different duration of attack simulated on a Splash benchmark suite [18].

But if those instructions update the LLC then the changes become permanent. Thus, the LLC write buffers only write back at the time of checkpoint. If they are full before the checkpoint the processor is halted and the checkpointing is performed. The frequency of checkpointing is dependent on the size of the write buffer. This issue can also be solved by adding a volatile bit to each cache line which when set will indicate that the contents are speculative and the volatile data is committed during the checkpoint [10]. After the data is committed the volatile bits are cleared.

## IV. Simulation Results

The proposed bypass architecture is evaluated on a 4-core Alpha processor in gem5 [17]. The configuration of the processor cores is provided in Table 1. The gem5 code is modified to implement: (a) variable read and write latency for STTRAM LLC; (b) an attack signal is added which is turned ON dynamically to mimic the actual attack signal from the sensors; and, (c) bypassing of LLC is implemented by modifying the cache access method to force a miss when the attack signal is high. The simulations are performed on a wide range of SPLASH benchmarks suite [18]. Two scenarios have been considered to calculate the loss in performance and energy due to the proposed bypass architecture during attack: (a) the attack persists for the entire duration of the simulation; and, (b) the attack is withdrawn after ~25%, ~50% and ~75% completion of each benchmark. There is no bypass for the rest of the time.

Fig. 7(a) shows the instruction per cycle (IPC) for the two scenarios compared with the normal execution without an attack. In case of 100% attack the system behaves as if there is no LLC and thus the performance degrades by an average of 27% (maximum 40%). For 50% attack the performance degradation is 8% (average) and 15% (max) but in both cases the system continues computation during the attack. Fig. 7(b) shows the energy overhead of bypassing. In case of 100% attack the energy overhead is 2% (average) and 5%(max). The energy is calculated using the multicore power simulator McPAT[19] with modified CACTI [20].

## V. Conclusions

Applicability of emerging technologies such as STTRAM in in memory hierarchy faces security challenges due to possibility of low-cost non-invasive tampering using external AC/DC magnetic field or temperature in order to launch denial-of-service attacks. We proposed three low-overhead solutions to mitigate these attacks: stalling, cache bypassing and system level checkpointing with bypassing. In case of gradually ramping attack we bypass the LLC and continue computation. For sudden attack we restore the processor to the last checkpointed state and continue computation with bypassing. The simulation results show an average of 27% (2%) overhead in IPC (energy) with the proposed bypass architecture for an attack lasting for the entire duration of execution.

## References

[1] Kryder, Mark H., et al. "After hard drives—what comes next?."Magnetics, IEEE Transactions on 45.10 (2009): 3406-3413.
[2] Jang, Jae-won, et al. "Self-Correcting STTRAM under Magnetic Field Attacks", DAC, 2015.
[3] Bi, Xiuyuan, et al. "Analysis and optimization of thermal effect on STT-RAM Based 3-D stacked cache design.", ISVLSI, 2012.
[4] Gupta, Swastik, et.al. "Adaptive cache bypassing for inclusive last level caches.",IPDPS, 2013.
[5] Gao, Hongliang, et al. "A dueling segmented LRU replacement algorithm with adaptive bypassing.", JWAC, 2010.
[6] Atkins, Mark. "Performance and the i860 microprocessor." Micro, IEEE 11, no. 5 (1991): 24-27.
[7] Wang, Jue, et al. "OAP: an obstruction-aware cache management policy for STT-RAM last-level caches.", DATE, 2013.
[8] Huangfu, Yijie, et al. "Real-Time GPU Computing: Cache or No Cache?.", ISORC, 2015.
[9] Kothari, L, et al. "Architecture of a checkpointing microprocessor that incorporates nanomagnetic devices." TComp, 2007.
[10] Shyam, Smitha, et al. "Ultra low-cost defect protection for microprocessor pipelines." In *ACM Sigplan Notices*, 2006.
[11] Martínez, José F., et al. "Cherry: Checkpointed early resource recycling in out-of-order microprocessors.", MICRO, 2002.
[12] Schulz, Martin, et al. "Implementation and evaluation of a scalable application-level checkpoint-recovery scheme for MPI programs.", SC, 2004.
[13] Zhang, Jianwei, et al. "Identification of transverse spin currents in noncollinear magnetic structures." Physical review letters, 2004.
[14] Schulz, Martin, et al. "Implementation and evaluation of a scalable application-level checkpoint-recovery scheme for MPI programs.", SC, 2004.
[15] Bronevetsky, Greg, et al. "Automated application-level checkpointing of MPI programs." ACM Sigplan Notices, 2003.
[16] http://www.google.com/patents/US5097532
[17] Gem5, http://www.gem5.org.
[18] Splash, http://kbarr.net/splash2
[19] McPAT, http://www.hpl.hp.com/research/mcpat
[20] CACTI, http://www.hpl.hp.com/research/cacti/.